\documentclass[11pt]{article}
\pdftrailerid{}             
\usepackage[letterpaper,margin=1in]{geometry}
\usepackage{setspace}
\usepackage[T1]{fontenc}
\usepackage{amsmath}
\usepackage{newtxtext,newtxmath}   
\usepackage{microtype}

\usepackage{graphicx}
\usepackage[font=small,labelfont=bf,skip=6pt,tableposition=top]{caption}
\usepackage[font=small]{subcaption}
\usepackage{booktabs}
\usepackage{titlesec}
\usepackage{array}
\usepackage{tabularx}
\usepackage[table]{xcolor}         
\usepackage{pifont}                
\usepackage{enumitem}
\usepackage[section]{placeins}     
\usepackage[title,titletoc]{appendix}

\definecolor{okgreen}{RGB}{0,128,0}
\definecolor{badred}{RGB}{190,30,30}
\newcommand{\yes}{\textcolor{okgreen}{\ding{51}}}
\newcommand{\no}{\textcolor{badred}{\ding{55}}}

\usepackage[natbibapa]{apacite}    

\usepackage{hyperref}
\hypersetup{
  colorlinks=true,
  linkcolor=blue!50!black,
  citecolor=blue!50!black,
  urlcolor=blue!50!black,
  pdftitle={Analyzing Memory Effects in Large Language Models through the Lens of Cognitive Psychology},
  pdfauthor={},
  pdfsubject={},
  pdfkeywords={},
  pdfcreator={},
  pdfproducer={}
}

\title{Analyzing Memory Effects in Large Language Models\\
       through the Lens of Cognitive Psychology}
\author{%
  \begin{tabular}[t]{c}
    Zhaoyang Cao\\
    {\small Data Lab, Dept. of Electrical}\\
    {\small Engineering and Computer Science}\\
    {\small Syracuse University}\\
    {\small zycao@data.syr.edu}
  \end{tabular}
  \hspace{0.6cm}
  \begin{tabular}[t]{c}
    Lael Schooler\\
    {\small Department of Psychology}\\
    {\small Syracuse University}\\
    {\small lschoole@syr.edu}
  \end{tabular}
  \hspace{0.6cm}
  \begin{tabular}[t]{c}
    Reza Zafarani\\
    {\small Data Lab, Dept. of Electrical}\\
    {\small Engineering and Computer Science}\\
    {\small Syracuse University}\\
    {\small reza@data.syr.edu}
  \end{tabular}
}
\date{}

\begin{document}
\maketitle

\begin{abstract}
Memory, a fundamental component of human cognition, exhibits adaptive yet fallible characteristics as illustrated by Schacter's memory ``sins.'' These cognitive phenomena have been studied extensively in psychology and neuroscience. Importantly, these ``sins'' are not merely flaws, but reflect tradeoffs inherent in an efficient and adaptive memory system. The extent to which artificial systems, specifically Large Language Models (LLMs), emulate these cognitive phenomena remains underexplored. This study uses human memory research as a lens for understanding LLMs, systematically investigating memory effects in state-of-the-art LLMs using established experimental paradigms. We evaluate seven key memory phenomena, comparing human behavior to LLM performance. Both people and models remember less when overloaded with information (list length effect) and remember better with repeated exposure (list strength effect). They also show similar difficulties when retrieving overlapping information, where storing too many similar facts leads to interference (fan effect). To further explore the mechanism underlying this interference, we conduct an extended-range fan experiment and analyze token-level response probabilities. As associative fan increases, target hit rates remain near ceiling, whereas false alarm rates for recombined foils increase substantially. This asymmetry suggests that the observed effect is driven primarily by familiarity-like evidence for recombined probes and contextual competition, rather than by a simple loss of the studied facts. Like humans, LLMs are susceptible to falsely ``remembering'' words that were never shown but are related to others (false memories), and they can apply prior learning to new, related situations (cross-domain generalization). However, LLMs differ in two key ways: they do not exhibit human-like U-shaped serial position effects at the list lengths tested and are more robust to nonword or semantically meaningless material.
These results reveal both alignments and divergences in the reconstructive nature of memory in LLMs and humans. The findings help clarify how memory-like behavior in LLMs echoes core features of human cognition, while also highlighting behavioral differences that lead to distinct patterns of error and success.
\end{abstract}


\bigskip

\section{Introduction}
As the basis for learning, decision-making, and identity development, memory is an essential part of human cognition \citep{lynch1991memory,anderson2023environmental,baddeley2013essentials}. Yet, despite its vital role in daily life, memory is not always infallible when it comes to our experiences \citep{huemer1999problem,singer2002tell}. Instead, it is a constructive process, vulnerable to illusion and error \citep{winograd1998individual,goodman2007memory,roediger1996memory,anderson1989human}. Research in cognitive psychology has illuminated various complexities of memory, including its notable ``malfunctions.'' Notably, \citet{schacter1999seven,schacter2002seven,schacter2003seven,schacter2022seven} offers a taxonomy of seven memory ``sins,'' which include \textit{transience}, \textit{absent-mindedness}, \textit{blocking}, \textit{misattribution}, \textit{suggestibility}, \textit{bias}, and \textit{persistence}. Each of the sins reflects the adaptive qualities of human memory while also offering significant insights into its underlying constructive character. Various related fields in psychology, such as cognitive psychology \citep{roediger1980memory}, social psychology \citep{blank2009remembering}, and clinical psychology \citep{squire1982neuropsychology}, along with cognitive neuroscience \citep{baddeley1997human,gabrieli1998cognitive,squire2011cognitive,rugg2003human}, have conducted extensive studies on these memory errors. These errors are interpreted as adaptive byproducts, highlighting their evolutionary significance in human cognition and providing a framework not only to understand how memory malfunctions but also how it functions in complex environments.

Although human memory has been extensively studied, the emergence of Large Language Models (LLMs) provides a novel domain in which to apply and test the cognitive toolkit developed in memory research. Recent LLMs like GPT \citep{achiam2023gpt} and DeepSeek \citep{guo2025deepseek} incorporate mechanisms that resemble large-scale associative memories, retrieving knowledge through contextual search and reconstructing information, perhaps in ways that are functionally similar to how human memory operates. These models, powered by the transformer architecture \citep{vaswani2017attention}, have achieved great success in various applications, including but not limited to language translation \citep{raffel2020exploring,donthi2025improving}, text summarization \citep{pu2023summarization,jin2024comprehensive,zhang2024comprehensive}, and even composing music and poetry \citep{yuan2024chatmusician,bechard2024pen,ding2024songcomposer}. These systems process information through layered parameterized transformations inspired by biological neural plasticity, yet the extent to which they mimic human cognitive processes remains under-explored \citep{niu2024large,ren2024large,tang2024humanlike}.

Benchmarks reveal that advanced LLMs like DeepSeek achieve superior performance on associative reasoning tasks compared to other state-of-the-art language models \citep{liu2024deepseek,liu2024deepseek1}. These tasks encompass multiple cognitive dimensions, including but not limited to (1) multi-subject multiple-choice evaluation, (2) language understanding and reasoning, and (3) reading comprehension, among others. The observed performance improvements suggest that architectural innovations in models may better approximate human-like associative reasoning mechanisms in cognitive systems. This raises fundamental questions: \textit{Do LLMs exhibit memory phenomena similar to those observed in human cognition?} Specifically, (1) Which human memory effects are replicated in LLMs? and (2) Where do LLMs diverge from human memory systems, potentially revealing their mechanistic limitations? Such analysis could establish whether LLM `memory' operates via principles similar to human cognition, and these differences may help us understand how associative memory works in LLMs and suggest ways to improve their performance.

Recently, \citet{binz2024centaur} proposed \textit{Centaur}, a computational model capable of predicting and simulating human behavior.
They developed Centaur by fine-tuning it on a large-scale dataset, ``\textit{Psych-101},'' which contains trial-by-trial human data from experimental paradigms including multi-armed bandits, decision-making, memory, supervised learning, Markov decision processes, and others. Their framework addresses early skepticism toward unified cognitive models by proposing a ``cognitive decathlon'' evaluation paradigm (a rigorous evaluation framework wherein sixteen experiments are used to test cognitive models, and their combined performance is evaluated), wherein Centaur outperformed established models across all of the experimental competitions. Centaur's success in simulating domain-general cognition highlights the promise of data-driven approaches to cognitive modeling. Nevertheless, while Centaur focuses primarily on broad behavioral alignment, our research specifically investigates how large language models replicate or diverge from fundamental human memory phenomena.

Unlike the aforementioned research that employs LLMs to model human cognitive processes, our study uses human memory research as a lens for understanding LLMs. Human memory has been the subject of meticulous research for more than 140 years, beginning with \citep{ebbinghaus1964memory}, and this research has produced numerous computational models of memory \citep{anderson1998integrated,anderson1989human,gillund1984retrieval,hintzman1984minerva,murdock1982theory,raaijmakers1981search,shiffrin1997model}; for a review, see \citep{kahana2020computational}. These models provide a framework and guide to understanding the functioning of LLMs. Rather than infer the internal mechanisms of LLMs directly, for example, by inspecting model weights, we adopt human memory paradigms as a behavioral diagnostic lens. Our goal is to identify where behavioral regularities align or diverge under controlled memory-like tasks.

Our investigation evaluates experimental paradigms from human memory research---primarily associative recognition studies of the fan effect \citep{anderson1974retrieval,anderson1999fan} and the Deese-Roediger-McDermott (DRM) false memory \citep{roediger1995creating}---to evaluate memory-like behavior in LLMs. We evaluate seven critical phenomena: list length effect (scaling performance with increasing memory load), list strength effect (interference in repetitive retrieval), fan effect (associative interference), lexicality manipulation (meaningful words vs. nonsense strings), serial position effect (primacy/recency patterns in sequential memory), DRM-style false memories (false memory in a list learning paradigm), and cross-domain generalization ability (transfer of schema-consistent details). Our major contributions are summarized as follows.
\begin{itemize}[leftmargin=*, itemsep=2pt, topsep=4pt]
\item We conduct a systematic investigation of human-like memory effects (including list length/strength effects, fan effects, lexicality manipulation, serial position effect, and DRM-style false memories) across several state-of-the-art LLMs, providing the information for evaluating progress toward human-level memory capabilities in LLMs.
\item We demonstrate that LLMs exhibit similar characteristics of human memory effects: (i) human-like patterns in associative interference---the fan effect; (ii) similar sensitivity to list length and list strength variations; (iii) parallel susceptibility to DRM-style false memories; and (iv) comparable generalization ability. While LLMs mirror these qualitative patterns, they are more accurate and less affected by interference, even as their pattern of errors remains similar to those of human cognition.
\item Our experiments reveal where LLM memory systems fundamentally differ: (i) superior robustness to semantically anomalous content (lexicality manipulation); and (ii) less influence of the order in which information is encoded.

\item Beyond establishing behavioral correspondence, we extend the fan manipulation from the classic-range fan experiment to seven levels from 1 to 20 and analyze log-odds. Target hit rates remain near ceiling, while false alarms to recombined foils increase with fan. Foil evidence shifts toward and across the old/new criterion and becomes more flattened, indicating that the manipulation acts primarily on the foil side of the recognition decision, 
a pattern that aligns closely with computational accounts of memory such as rational analysis and ACT-R.\footnote{Data and analysis code are available at \url{https://anonymous.4open.science/r/LLM_CognitivePsychology-731E/README.md}.}
\end{itemize}

\section{Theory Justification and Explanation}
In research on memory and human cognition, various effects have been identified to describe how humans process and recall information. More specifically, memory retrieval can be broadly categorized into two aspects: recall and recognition. Recall refers to the active retrieval of information from memory in response to cues, reflecting a generative process. In contrast, recognition entails the identification of information as previously encountered when presented with a cue. In this section, we first introduce the memory effects involved in the experiments in this paper. We will then present research that looks for patterns in large language models that parallel human recognition. Throughout this paper, we use memory retrieval as an umbrella term encompassing recognition, free recall, and cued recall. These labels refer to task demands rather than to distinct cognitive processes. Importantly, responses in all of these tasks may reflect reconstructive processes, drawing on stored memories as well as biasing or contextual information available at retrieval. We reserve the term recall for cases in which free recall is specifically intended.
\newline\noindent $\blacktriangleright$ The \textbf{\textit{list length effect}} refers to the decline in memory as the number of studied items increases. In free recall, the effect is reliable \citep{murdock1962serial}. In recognition, it has been debated: early studies reported lower accuracy for longer lists \citep{gronlund1994list}, but when retention interval, attention during study, and contextual reinstatement at test are controlled, the effect becomes small or disappears \citep{dennis2001context,kinnell2011list}. Its absence was a key motivation for context-noise models. In these models, recognition errors come mainly from the other contexts in which an item has appeared \citep{dennis2001context}. This makes the list length effect diagnostic for LLMs: a recognition decline with list length would be a divergence from controlled human data. 
\newline\noindent $\blacktriangleright$ The \textbf{\textit{list strength effect}}, systematically investigated by Ratcliff, Clark, and Shiffrin, represents a fundamental characteristic of competitive memory retrieval. This phenomenon manifests in two distinct ways: in free recall, strengthening certain items in a memorized list (typically through repetition or extended study time) harms the remaining list items; in recognition, it appears as a missing or negative effect \citep{ratcliff1990list,shiffrin1990list}. 
\newline\noindent $\blacktriangleright$ The \textbf{\textit{lexicality manipulation}} tests whether meaningful material is remembered better than meaningless material. In serial recall, words are recalled better than nonwords \citep{hulme1991memory}. In associative recognition, \citet{xu2007modeling} varied the type of stimuli forming the pairs, including words and pseudowords, and found a mirror effect: pairs of better-recognized stimuli produced both higher hit rates and lower false-alarm rates. They attributed this to words being encoded more extensively than nonverbal stimuli. 
\newline\noindent $\blacktriangleright$ The \textbf{\textit{serial position effect}} refers to better memory for items at the beginning (primacy) and end (recency) of a list than for items in the middle. It was established in free recall \citep{murdock1962serial,glanzer1966two}; in recognition, the curve is usually flatter. Analogous position biases have been reported for LLMs processing long contexts, where information in the middle of the input is used less well \citep{liu2024lost,hsieh2024found,veseli2025positional}.
\newline\noindent $\blacktriangleright$ The \textbf{\textit{Deese-Roediger-McDermott paradigm}} (known as the DRM effect) illustrates how people recall related but non-presented words in a list, revealing the impact of semantic activation on false memory. When participants study lists of strongly related words (e.g., thread, pin, eye, sewing for the critical lure `needle'; or bed, rest, tired, dream for `sleep'), they often falsely recall or recognize the non-presented lure word \citep{newstead1998false,cortese2008activation,coane2021manipulations}. This occurs because semantically related words activate the lure concept in memory, which causes it to be incorrectly recalled alongside elements that are actually presented \citep{roediger1995creating}. Additionally, the DRM effect is often used as a model laboratory task for the formation of false memories in people. 
\newline\noindent $\blacktriangleright$ The \textbf{\textit{generalization ability}} refers to the tendency of memory to extend what was learned from specific items to new items that share their meaning. It reflects a tradeoff between two kinds of memory traces: verbatim traces preserve the surface details of what was experienced, whereas gist traces preserve its meaning \citep{brainerd2008developmental}. When verbatim traces are weak, people reconstruct what they experienced from gist and from prior knowledge, keeping the core meaning while filling in details that are consistent with it \citep{bartlett1995remembering}. Generalization thus allows memory to be applied flexibly to new but related situations, at the cost of accepting items that fit the gist of what was studied but were never presented.
\newline\noindent $\blacktriangleright$ Lastly, the \textbf{\textit{fan effect}} proposed by Anderson describes how an increase in associations to an item leads to slower retrieval times for information associated with the item \citep{anderson1999fan}. It was critical to the development of ACT-R \citep{anderson1974retrieval,anderson1999fan} and is a well-studied case of associative recognition \citep{clark1996global,yonelinas1997recognition,rotello2000recall}. That makes it the natural place to ask not merely whether a language model reproduces a human pattern, but whether it does so in a form an existing memory model would recognize. Formally, a fan refers to the number of associations of a concept in a semantic network. \citet{schneider2012modeling} examined the time course of associative recognition with a speed--accuracy tradeoff procedure, in which accuracy is measured as a function of the time allowed for retrieval. In their experiment, higher fan slowed the rate at which accuracy rose toward its asymptote but left the asymptote itself unchanged. A lower asymptote for high-fan pairs appears instead in the data of \citet{wickelgren1977associative}, which Schneider and Anderson also modeled. The fan effect is particularly useful for moving beyond a behavioral comparison because cognitive models make explicit predictions about how associative competition affects retrieval. 

The rational analysis underlying ACT-R's account of declarative memory regards memory as adapted to predicting which information will be needed in a given context \citep{anderson1989human,schooler2017adaptive}. This provides a conceptual linkage to the next-token prediction objective used in language-model pretraining. In ACT-R, each memory is stored as a \emph{chunk}, a small unit of declarative knowledge. In the experiments considered here, a memory is a basic proposition, such as a studied person--location fact, or, in the DRM experiments, a single word. The activation of chunk $i$ can be expressed as
\begin{equation}
    A_i = B_i + \sum_j W_j S_{ji},
    \qquad
    S_{ji} = S - \ln(\mathrm{fan}_j),
    \label{eq:actr_fan}
\end{equation}
where $B_i$ denotes base-level activation, $W_j$ is the source activation allocated to cue $j$ (source activation is divided equally among the $J$ cues and sums to a constant $W$, so $W_j = W/J$), $S$ is the maximum associative strength, and $S_{ji}$ is the associative strength from cue $j$ to chunk $i$ \citep{schneider2012modeling}. Here, ACT-R motivates the evidence measure rather than serving as a fitted model of LLM behavior. Two parallels from computer science, specifically from analyses of the transformer architecture \citep{vaswani2017attention}, make this mapping concrete. In a transformer, self-attention is the operation by which the representation of each token is updated from the representations of the earlier tokens in the context. First, self-attention can be written as content-addressable retrieval: a query is compared with the keys computed for every earlier token and returns a similarity-weighted mixture of their values, which is formally equivalent to one update of a modern Hopfield network \citep{ramsauer2021hopfield} and analyzable as an associative memory \citep{bietti2023birth}. Second, induction heads retrieve from an earlier span that matches the current tokens and copy what followed it \citep{olsson2022context}. Because the studied facts in our tasks exist only in the context, we treat the tokens of one studied fact as the counterpart of an ACT-R chunk, and the probe's overlap with those tokens as the counterpart of the retrieval cues $j$ in Equation~\ref{eq:actr_fan}. The experiments use an old/new recognition test. The model first studies a list of facts or words and is then shown test probes one at a time: studied items are ``old,'' and recombined pairs or unstudied words are ``new.'' The model answers yes (old) or no (new) to each probe. Since our experiments do not directly measure an activation comparable to that of an ACT-R chunk, we instead define a graded old/new evidence score from the model's output probabilities. For a probe $q$,
\begin{equation}
    a(q) = \log \frac{P(\mathrm{old}\mid q)}{P(\mathrm{new}\mid q)}.
    \label{eq:llm_old_activation}
\end{equation}
Positive values refer to an ``old'' response, negative values refer to ``new,'' and zero is the decision criterion. This log-odds score is an activation-like decision variable, but it is not assumed to be mechanistically identical to ACT-R activation. We use it to examine whether increasing fan changes the evidence assigned to studied targets, recombined foils, or both.

Collectively, these effects serve as benchmarks to assess the alignment of memory dynamics in human cognition with large language models.

\section{Dataset Modification}
Our experimental framework leverages two datasets to systematically evaluate memory phenomena in large language models. The first dataset, ``Person Location Lists'' (see Supporting Information Table S1; hereafter, the \textit{\underline{person--location dataset}}), builds upon person-location pairs from \citet{schneider2012modeling}, which we augmented with additional words to enable comprehensive testing of the \textbf{\textit{list length/strength effects, fan effect}}, and \textbf{\textit{lexicality manipulation}}. The second dataset, ``Target Words and Associates Lists'' (see Supporting Information Table S2; hereafter, the \textit{\underline{lure--associate dataset}}), comprises carefully selected word association lists derived from \citet{roediger1995creating}, specifically chosen to assess \textbf{\textit{DRM-style false memory formation, serial position effect}}, and \textbf{\textit{generalization ability}}.

\subsection{Person--Location Dataset}\label{dataset1}
The dataset is organized into four controlled sub-experiments (Experiments 1--4) that evaluate fundamental memory effects: Experiment 1 examines memory capacity via list length variations. Experiment 2 manipulates item strength to examine the list-strength effect by repeating a selected study fact one or four additional times. Experiment 3 targets associative interference through the fan effect paradigm, and Experiment 4 manipulates lexicality. Specifically, we prepared (I) a study list, (II) a foil list, and (III) a test list. The study list contains a series of facts, each following the form ``\texttt{The <person> is in the <location>.}'' These were generated from sets of 40 persons and 40 locations. The word lengths for persons and locations ranged from 3 to 12 characters (mean = 6.85, sd = 1.93) and from 4 to 10 (mean = 6.33, sd = 1.83), respectively. The size of the study list varied according to each sub-experiment. The foil list included non-studied facts created by rearranging persons and locations while maintaining the \textit{fan status} (word frequency) of the study list. The test list covered both the study and foil facts, converting their facts into question format using the form ``\texttt{Is the <person> in the <location>?}'' The test list contained every studied fact (target) and an equal number of recombined foils. Probe order was randomized separately for each run with a recorded seed, so that a probe's position carried no information about the correct response.

\subsubsection{Synthetic Object--Location Fan Datasets}\label{sec:synthetic}
To complement the person-location dataset with a more controlled test of associative interference, we developed a stimulus generator for synthetic object--location memory tasks. It creates novel identifiers and matched study and recognition lists while controlling associative fan and presentation frequency. The extended-range fan experiment used seven fan levels ($1,2,3,4,7,12,$ and $20$). At each level, 20 object identifiers and 20 location identifiers were generated; a fan-$N$ block contained $20N$ studied pairs. Across the seven levels, each replication therefore contained 980 studied pairs and 980 recombined foils that preserved the constituent fan values, for 1,960 test probes. Since the identifiers are generated rather than drawn from semantic categories, the resulting stimuli reduce the possibility that recognition judgments are driven by pre-existing semantic associations or memorized training examples.
Five independent replications were conducted, each with a new set of object and location identifiers, studied pairs, and test probes. The pooled analysis therefore comprised 9,800 probes. Thus, the reported effect does not depend on one particular assignment of words to fan conditions or on one fixed study/test list. Studied pairs and test probes were intermixed across fan levels in random order, so fan was not confounded with study or test position. The stimulus generator and the corresponding datasets are released publicly.

\begin{table}[tp]
  \centering
  \caption{Confusion matrix in the memory experiment.}
  \label{confusionmatrix}
  \renewcommand{\arraystretch}{1.3}
  \begin{tabular}{lcc}
    \toprule
    & \textbf{Predicted Positive (``old'')} & \textbf{Predicted Negative (``new'')} \\
    \midrule
    \textbf{Actual Positive (target)} & TP & FN \\
    \textbf{Actual Negative (foil)}   & FP & TN \\
    \bottomrule
  \end{tabular}\vspace{-2mm}
\end{table}

\subsection{Lure--Associate Dataset}\label{dataset2}
The dataset is organized into three additional controlled sub-experiments (Experiments 5--7) that evaluate distinct memory phenomena: Experiment 5 examines recall patterns via serial position effects, Experiment 6 investigates false memory formation through the Deese-Roediger-McDermott paradigm, and Experiment 7 assesses memory flexibility through cross-domain generalization ability. Specifically, we selected 12 target words for the experiment, and for each target word, 15 associated words were selected for the list. The words in the list were ordered based on the strength of their association, with the most strongly associated words appearing first. The selection and ordering of target words and their associates followed the lists provided by \citet{roediger1995creating}.

\section{Materials and Methods}
All experiments in this study are mainly conducted using an online large language model, specifically the GPT-4 architecture. More precisely, we utilize the \texttt{gpt-4.1} variant provided through the OpenAI API. The subsequent sections will systematically detail the evaluation metrics, procedures, and designs of experiments.

\subsection*{Evaluation Metrics}
Performance was assessed by recording recognition accuracy, false alarm rate, hit rate, and standard deviation. According to the confusion matrix presented in Table~\ref{confusionmatrix}, these evaluation metrics are defined mathematically as follows:
\begin{align}
\text{Recognition accuracy} &= \frac{\mathrm{TP} + \mathrm{TN}}{\mathrm{TP} + \mathrm{TN} + \mathrm{FN} + \mathrm{FP}}, \\
\text{False-alarm rate}     &= \frac{\mathrm{FP}}{\mathrm{FP} + \mathrm{TN}}, \label{equation3} \\
\text{Hit rate}             &= \frac{\mathrm{TP}}{\mathrm{TP} + \mathrm{FN}}. \label{equation4}
\end{align}

A core quantitative measure of recognition memory sensitivity is $d'$, derived from signal detection theory \citep{macmillan1985detection}. It quantifies how well a model can discriminate between \textit{targets} (studied items) and \textit{foils} (unstudied items), independent of response bias. The formula of $d'$ is shown in Equation~\ref{equation1} for our LLM experiments,
\begin{equation}\label{equation1}
d' = \Phi^{-1}(H) - \Phi^{-1}(F),
\end{equation}
where $H$ and $F$ denote the hit and false-alarm rates (Equations~\ref{equation4} and~\ref{equation3}), respectively, and $\Phi^{-1}$ is the inverse of the cumulative normal distribution. A higher $d'$ value indicates superior discriminability---meaning the system reliably distinguishes studied from unstudied pairs---while a $d'$ close to zero reflects near-chance performance and weak memory sensitivity.

For the extended-range fan experiment, we analyzed the model's token-level evidence in addition to its binary recognition response. For each probe $q$, we computed the old/new log-odds score $a(q)$ defined in Equation~\ref{eq:llm_old_activation} from the model's output token probabilities. A probe was scored as ``old'' when $a(q)>0$. We analyzed this score separately for targets and foils at each fan level to determine how increasing fan affected the two evidence distributions. Because the two distributions need not have equal variance, the extended-range experiment also examined per-fan z-ROC slopes and the associated unequal-variance sensitivity index $d_a$; equal-variance $d'$ was retained for comparison with the classic-range experiments.

\subsection*{Experimental Setup}
\noindent $\blacktriangleright$\textit{Prompts.} We interacted with the language models using task-specific prompts, one for each dataset (Supporting Information Tables~S3 and~S4). The format and content of these prompts were important for performing the recognition memory task accurately. We determined the final prompt design by various approaches, including interacting with the language models to get recommendations for optimal prompt structure for a recognition memory task. For the two open-weight models, we asked each model to revise the GPT prompt for dataset~1 to suit its own prompt requirements (Table~S5), which yielded the prompts used for Llama~3 and Mistral (Tables~S6 and~S7); for dataset~2, the original prompt was used for all models.
\newline\noindent $\blacktriangleright$\textit{Dataset I: person--location dataset.} We tasked large language models with memorizing facts during a study phase. Subsequently, a recognition task was performed during the test phase, where the language models were told to distinguish between targets (studied facts) and foils (non-studied facts, which were rearranged items derived from the studied facts). All studied facts and foils were presented in question format in the test phase.

We distinguish two fan experiments. The \emph{classic-range fan experiment} compared fan 1 and fan 2 using person-location pairs. The \emph{extended-range fan experiment} used the synthetic object--location dataset described in Section~\ref{sec:synthetic} and examined seven fan levels: 1, 2, 3, 4, 7, 12, and 20. The classic-range fan experiment used three fan conditions, referred to as Groups 1--3 throughout the paper. Here, fan denotes the number of studied associations involving a person or location. For the baseline 32-pair study lists, Group 1 (fan value = 1) contained 32 pairs involving 32 distinct persons and 32 distinct locations, each appearing once (Supporting Information Table~S8). Group 2 (fan value = 2) contained 32 pairs involving 16 distinct persons and 16 distinct locations, each appearing twice (Table~S9). Group 3 (mixed fan 1 and fan 2) contained 16 fan-1 pairs and 16 fan-2 pairs. Starting from these lists, we lengthened the study list (Tables~S10--S17 for fan~1 and Tables~S18--S21 for fan~2), repeated a selected study fact (Tables~S22--S25), and replaced persons, locations, or both with nonsense strings (Tables~S26--S28); these manipulations are detailed in Appendix~\ref{appendix_a}.

The outputs from the language models were then collected to assess recognition performance. A response was scored correct when the model answered ``yes'' to a target or ``no'' to a foil; scoring used the answer key generated together with each randomized order. The final evaluation results were obtained by comparing the recognition results with the ground truth.
\newline\noindent $\blacktriangleright$\textit{Dataset II: lure--associate dataset.} In this experiment, language models were told to act as participants in a memory experiment and were first tasked with memorizing a study list of words during the study phase. The list under the DRM experiment comprised 120 words from eight associate lists (Supporting Information Table~S29). In the delayed condition, the models then solved a short series of math problems. In the subsequent test phase, the models were presented with a recognition task. The detailed design of all experiments is provided in Appendix~\ref{appendix_a}.

\begin{figure}[tbp]
  \centering
  \includegraphics[width=\textwidth,height=0.5\textheight,keepaspectratio]{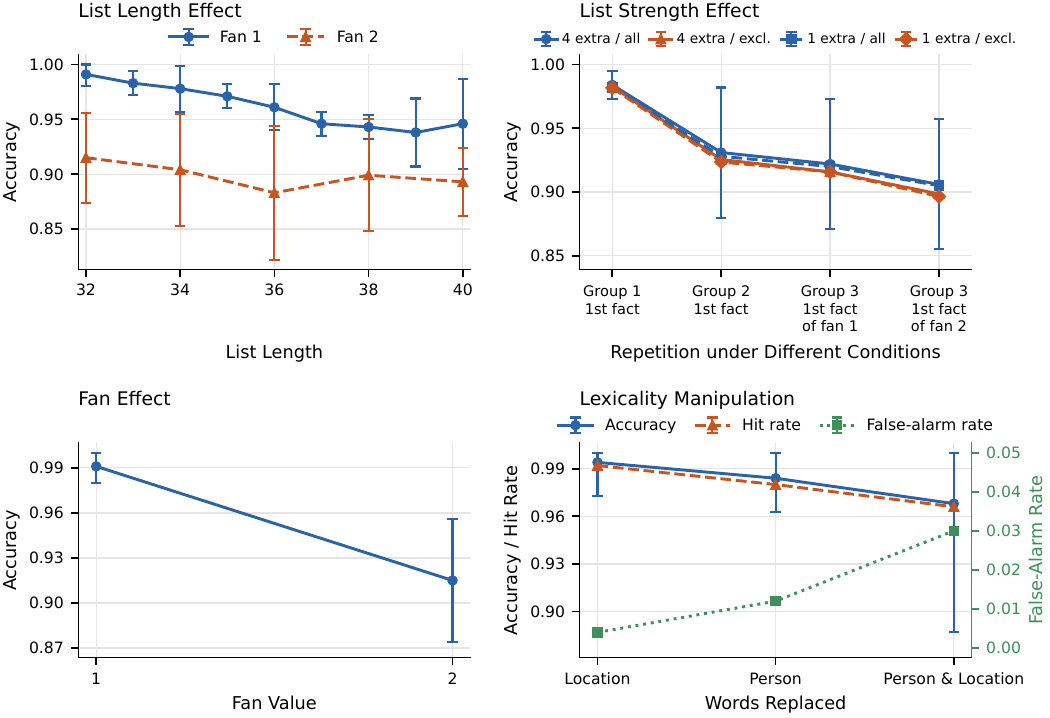}\vspace{-1mm}
  \caption{Performance of the online language model on the person--location dataset. Error bars show $\pm 1$ standard deviation across runs.
  \textbf{List length effect}: Study lists varied in size (32--40 person--location pairs), showing that recognition accuracy declines as list size increases.
  \textbf{List strength effect}: Strengthening a selected study fact through repetition was not accompanied by a significant reduction in recognition performance for the remaining non-repeated items.
  \textbf{Fan effect}: Fixing the study list size at 32 pairs and varying fan values demonstrated that higher associations to an item reduce recognition accuracy.
 \textbf{Lexicality manipulation}: Replacing person and location words with nonsense strings slightly lowered hit rates and raised false-alarm rates.}
  \label{result}
\end{figure}

\section{Results}
We conducted experiments with various large language models, including GPT-4, Mistral, and LLaMA, and observed consistent trends across these models. While some models demonstrated superior performance in specific memory-related tasks, none of the evaluated LLMs achieved perfect results across all memory experiments. Notably, GPT-4 exhibited the most robust performance among the tested models. Therefore, we primarily present the detailed results obtained from GPT as it consistently outperformed other models in our comparative analysis.

\begin{figure}[tbp]
  \centering
  \includegraphics[width=\textwidth,height=0.55\textheight,keepaspectratio]{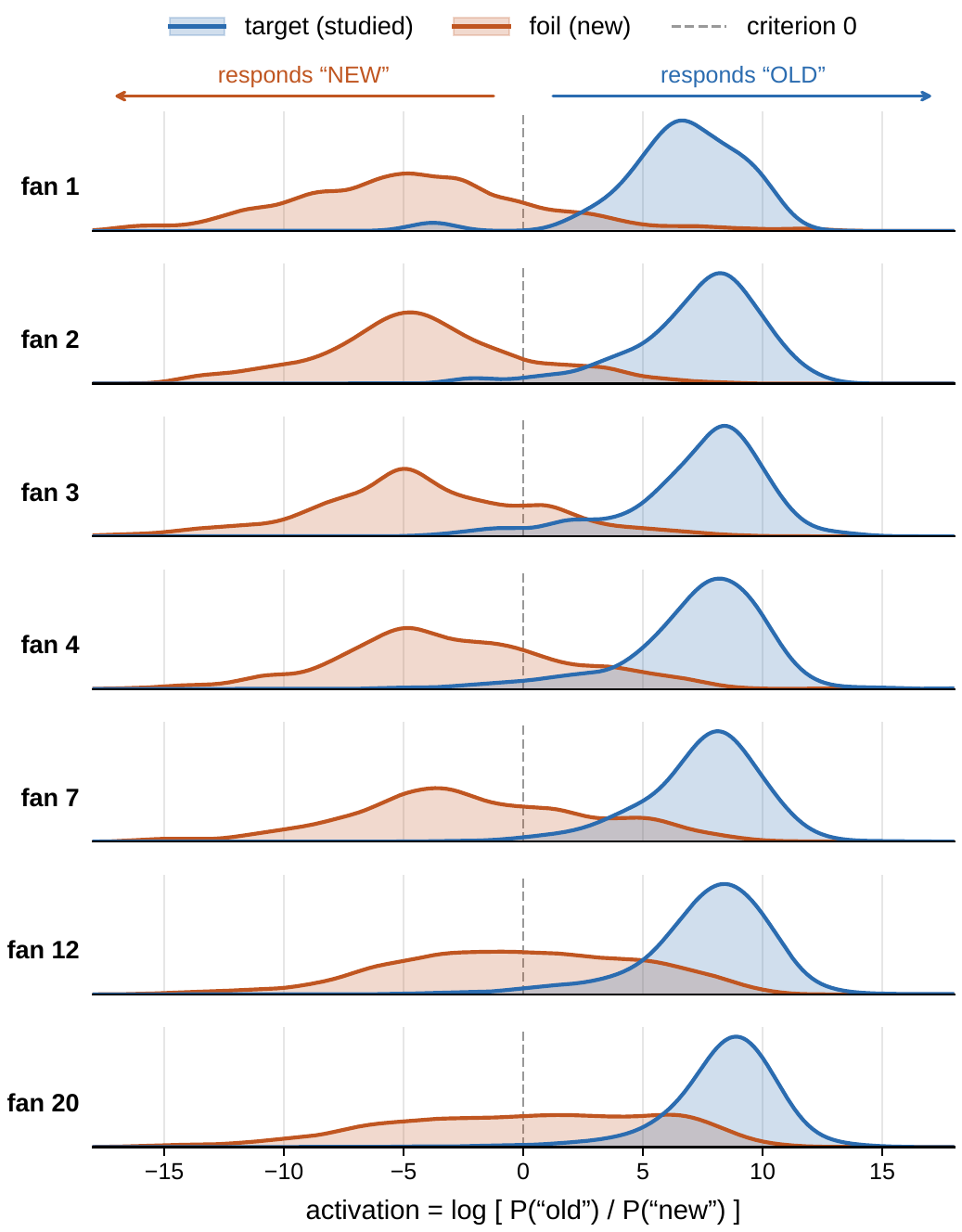}
  \caption{Old/new evidence distributions by associative fan for studied targets (blue) and recombined foils (orange). Evidence is defined as $a(q)=\log[P(\mathrm{old}\mid q)/P(\mathrm{new}\mid q)]$. The dashed vertical line marks the decision criterion $a(q)=0$; values to the right favor an ``old'' response. As fan increases, the foil distribution broadens and shifts toward the criterion while target evidence remains predominantly positive.}
  \label{fan_distributions}
\end{figure}

\subsection*{Person--Location Dataset}
The results on the person--location dataset are shown in Figure~\ref{result}, which leads us to the following conclusions. (1) Regarding the fan effect, when the fan value increases, the accuracy of the language model on the recognition test list decreases; for example, the value decreases from 0.991 to 0.915 when the list size is 32. Higher fan thus lowers recognition accuracy. (2) With respect to the list length effect, the recognition accuracy of the language model decreases as the list size (the number of studied facts) increases, even though there are some fluctuations. (3) For the list-strength manipulation, one study fact was presented one or four extra times. To test whether strengthening one item impairs memory for the others, we scored the test twice: over all test items, and with the repeated fact excluded, since the repeated fact itself should be recognized better. Accuracy with the repeated fact excluded was slightly lower than accuracy over the full test list, but increasing repetition did not substantially reduce recognition of the remaining items. (4) For the lexicality manipulation, replacing person words, location words, or both with nonsense strings reduced recognition accuracy only slightly (from 0.994 to 0.968). Hit rates fell slightly while false-alarm rates rose, the mirror pattern that \citet{xu2007modeling} observed for pseudowords in human associative recognition.

\subsubsection*{Analysis of the Extended-Range Fan Experiment}
The extended-range fan experiment localized the recognition decline to recombined foils. Scoring a probe as ``old'' when $a(q)>0$, the hit rate for studied pairs remained near ceiling across the full range, from $0.970$ at fan 1 to $0.988$ at fan 20. In contrast, the false-alarm rate increased from $0.169$ to $0.525$. The beginning of this trend was flat: fan 1 and fan 2 differed little, with the increase starting from fan 3. Therefore, the decline in the extended-range fan experiment was not primarily a failure to recognize studied pairs; it reflected an increasing tendency to classify new combinations of familiar constituents as ``old.''

\begin{figure}[tbp]
  \centering\vspace{-1em}
  \includegraphics[width=0.62\textwidth]{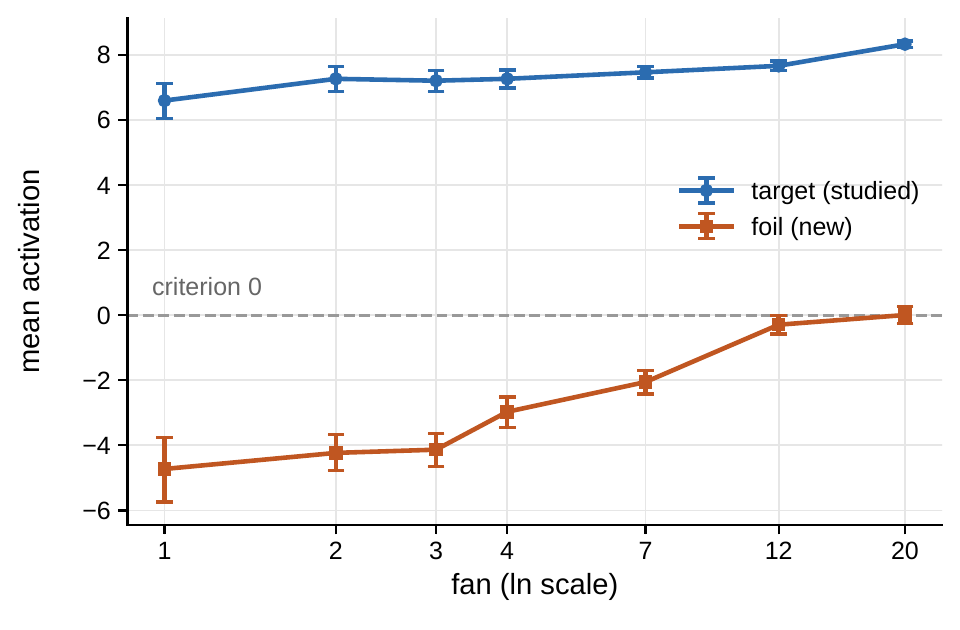}
  \caption{Mean old/new log-odds by fan for studied targets (blue) and foils (orange). Error bars show 95\% bootstrap confidence intervals; the dashed line marks $a(q)=0$.}
  \label{fan_mean_logodds}\vspace{-1em}
\end{figure}

Figure~\ref{fan_distributions} illustrates this asymmetry by showing full log-odds distributions separately for targets and foils at each fan level. A clear shift happens for foils: as fan increases, the foil distribution progressively moves rightward toward the decision criterion at zero. At low fan levels, most foils receive strongly negative evidence and fall on the ``new'' side of the criterion. With increasing fan, the foil distribution becomes flatter and shifts rightward, so an increasing portion approaches and eventually crosses the criterion. In contrast, the target distributions remain concentrated and similarly above the criterion across fan levels.

Figure~\ref{fan_mean_logodds} summarizes the evidence distributions by plotting their means across fan levels. The mean foil log-odds rises from approximately $-4.7$ at fan 1 to $0$ at fan 20, approaching the fixed old/new criterion. The target mean remains above zero, increasing from approximately $6.6$ to $8.3$.

The error bars in Figure~\ref{fan_mean_logodds} show uncertainty about the means, whereas Figure~\ref{fan_distributions} shows the spread of individual trials. The foil interquartile range increases from approximately $6.1$ at fan 1 to $8.8$ at fan 20, while the target interquartile range changes from approximately $3.0$ to $2.2$. Thus, higher fan places more foil evidence near or above the criterion without eliminating strongly positive evidence for studied pairs.

\begin{figure}[tbp]
  \centering
  \includegraphics[width=\textwidth,height=0.4\textheight,keepaspectratio]{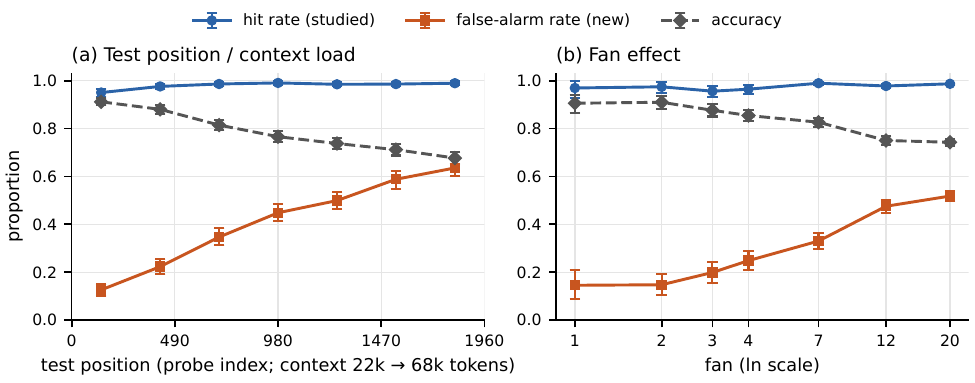}
  \caption{\textbf{(a)} Hit rate for studied targets (blue), false-alarm rate for recombined foils (orange), and overall accuracy (gray) as a function of test position (the probe's position in the test sequence), in bins of 280 consecutive probes. Because all probes were judged within one growing context, test position is equivalent to accumulated context length. \textbf{(b)} The same measures as a function of associative fan. In both panels, the decline in accuracy arises from rising false alarms, whereas hit rates remain near ceiling.}
  \label{fig:extended_effects}
\end{figure}

Since all probes of a replication were tested within one continuous context, each successive probe was judged against a longer context containing the study list and all preceding test items. Test position and accumulated context length are therefore confounded, and we treat them jointly as a measure of contextual load. Figure~\ref{fig:extended_effects}a shows that accuracy declined across test position, from $0.914$ in the first bin to $0.671$ in the last. This decline was carried by the foils: the false-alarm rate rose from $0.127$ to $0.649$, while the hit rate remained near ceiling and increased slightly, from $0.955$ to $0.990$. In signal-detection terms, sensitivity decreased ($d'$ from $2.83$ to $1.92$, with the log-linear correction). The drift did not reflect response perseveration, because the probability of an ``old'' response to a foil was nearly identical after an ``old'' versus a ``new'' response on the preceding trial.

Figure~\ref{fig:extended_effects}b demonstrates the fan manipulation as response proportions. Accuracy fell from $0.893$ at fan 1 to $0.740$ at fan 20, and the pattern again resided on the foil side: false alarms increased with fan while hits did not. Increasing contextual load and increasing associative fan thus produce the same signature. Neither weakens the evidence for studied pairs; both make recombinations of familiar constituents increasingly likely to be verified as old.

\subsection*{Lure--Associate Dataset}
Table~\ref{table_result_e2} illustrates the results on the lure--associate dataset. Whether it is an immediate recall or a later recall (after doing math problems), the hit rate (i.e., the proportion of studied words judged old) is always 1. In addition, for non-studied words, the false alarm rate of the language model is 0.053. This indicates that language models may incorrectly recognize words that were not present in the study list. The results are highly consistent with the DRM effect observed in humans. The proportion of critical lures judged old is 0.114, suggesting that language models are likely to generalize these lure words based on the associations in the study list, even though the lure words are not presented in the study list.

\begin{table}[bp]
  \centering
  \caption{Recognition results for studied items and critical lures in the lure--associate dataset.}\vspace{-1mm}
  \label{table_result_e2}
  \renewcommand{\arraystretch}{1.2}
  \begin{tabular}{lc}
    \toprule
    \textbf{Item type and condition} & \textbf{Proportion of ``old'' responses} \\
    \midrule
    Studied + immediate recall & 1.000 \\
    Studied + arithmetic delay & 1.000 \\
    Non-studied                & 0.053 \\
    Critical lure              & 0.114 \\
    \bottomrule
  \end{tabular}\vspace{-3mm}
\end{table}

\begin{figure}[tbp]
  \centering
  \includegraphics[width=\textwidth,height=0.5\textheight,keepaspectratio]{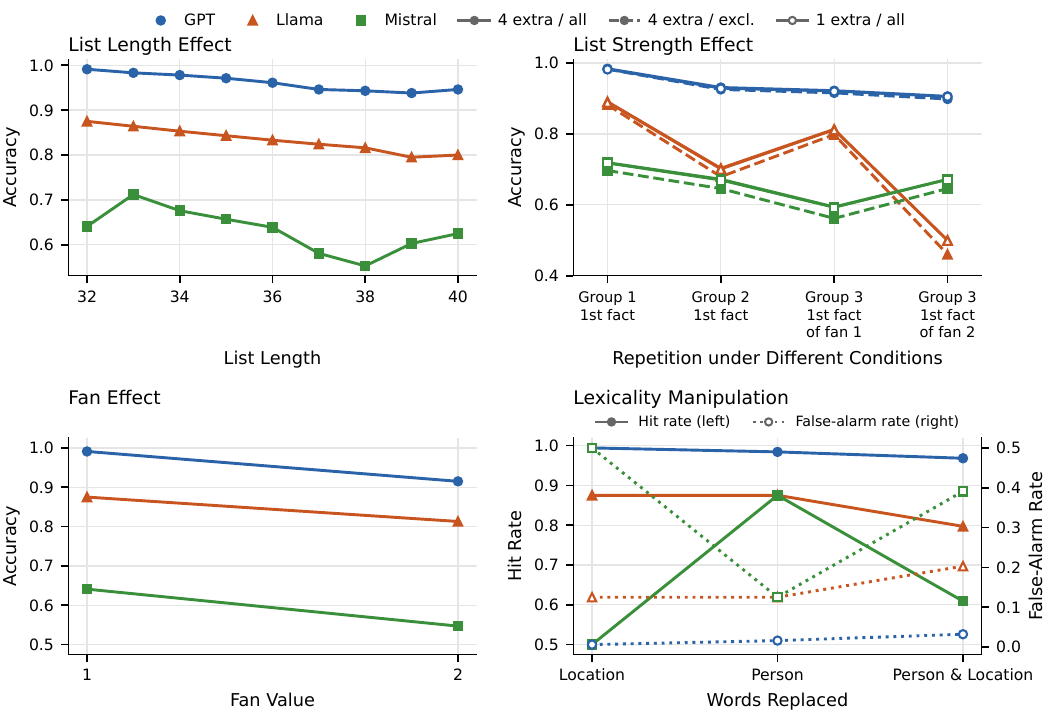}
\caption{Performance comparison between online and offline LLMs on the person--location dataset. The offline models (Mistral and Llama) show similar patterns across the four manipulations, with some fluctuations for Mistral in the list length effect and for Llama in the list strength effect. The online model consistently outperforms both offline models. Because decoding was deterministic, each offline condition was run once, so no error bars are shown.}
  \label{result2}
\end{figure}

In Experiment 5, the model recognized studied words perfectly at every serial position of the 12-word lists, so no serial position effect was observed. This differs from the human U-shaped serial position curve: participants in \citeauthor{roediger1995creating}'s experiment recalled words presented at the beginning and end of the study list more accurately than those in positions 4--8 \citep{roediger1995creating}. In the extended-range fan experiment, accuracy did decline across the test sequence (Figure~\ref{fig:extended_effects}a), but this decline is not a serial position effect. Studied pairs were presented in random order, hit rates remained near ceiling throughout, and the decline arose from rising false alarms as the accumulated context grew (see the analysis of the extended-range fan experiment above).

\subsection*{Comparison Analysis with Offline Models}
While much recent research focuses on online LLMs accessible via APIs, such as GPT-4, an important complementary direction involves understanding the capabilities of offline LLMs, which can be deployed locally without reliance on external servers. Unlike online models, which are often optimized for interaction, offline models provide greater reproducibility and control over the inference process. In this comparison with open-weight models, we extend our investigation of memory effects by applying frameworks from human memory research, aiming to evaluate whether similar cognitive patterns exist under constrained conditions. To enhance the possibility that these models can fully engage with our memory tasks, we selected state-of-the-art models with relatively large parameter counts. Specifically, we experimented with \texttt{Mistral-7B-Instruct-v0.3} and \texttt{LLaMA-3-8B} (prompts are given in Supporting Information Tables~S6--S7). Across all seven experiments spanning two datasets, we obtained complete results using each of these two offline models. A subset of the comparison results is presented in Figure~\ref{result2}.

\begin{table}[tbp]
  \centering
  \caption{Comparison of human and LLM performance on key memory effects.}
  \label{tab:memory_effects}
  \small
  \renewcommand{\arraystretch}{1.25}
  \rowcolors{2}{gray!10}{white}
  \begin{tabularx}{\textwidth}{>{\raggedright\arraybackslash}p{3.4cm}
                               >{\raggedright\arraybackslash}X
                               >{\raggedright\arraybackslash}X}
    \toprule
    \textbf{Memory Effect} & \textbf{Human Behavior} & \textbf{LLM Behavior} \\
    \midrule
    List Length Effect & Recall accuracy decreases as list length increases. & \yes\ Shows a parallel decrease in overall performance. \\
    List Strength Effect & Repetition of certain items leaves recognition of other items unaffected. & \yes\ Shows a parallel pattern, with recognition of weaker items not significantly affected by repetition of selected items. \\
    Fan Effect & Retrieval accuracy decreases with the number of associations. & \yes\ Shows the fan effect in memory tasks, showing less accurate outputs with increased fan. \\
    Lexicality Manipulation & Words are remembered better than nonwords, showing a mirror effect in associative recognition. & \no\ Superior robustness to semantically meaningless content. \\
    Serial Position Effect & Primacy and recency effects are observed. & \no\ No serial position effect in 12-item lists, where recognition was at ceiling; the decline across the test sequence in long lists reflects accumulated context, not serial position. \\
    DRM-style False Memories & Tends to falsely recall related but non-studied items. & \yes\ Exhibits lure responses consistent with the DRM paradigm, indicating similarity. \\
    Cross-domain Generalization & Flexible generalization with abstraction. & \yes\ Exhibits generalization but sometimes overfits specific examples, partially consistent. \\
    \bottomrule
  \end{tabularx}
\end{table}

We observed that for the four effects in the person--location dataset---list length, list strength, fan effect, and lexicality manipulation---the offline models exhibited a similar trend, despite some fluctuations (e.g., Mistral under the list length effect and Llama under the list strength effect). Nevertheless, online LLMs consistently outperformed offline LLMs. Additionally, because decoding was deterministic, repeated trials with identical prompts produced identical outputs, so each offline result reflects a single run per condition. Hence, Mistral shows limited performance in certain tasks in the lexicality conditions, such as when only locations are replaced or when both persons and locations are substituted. For the following effects in the lure--associate dataset, the offline models demonstrated perfect performance on the serial position effect as well, accurately recalling and recognizing words at all positions from the associated word lists. As for the DRM effect, offline models achieved a hit rate of 1, which is similar to GPT. These open-weight models also rejected every critical lure, so their discrimination was perfect: unlike human participants, they showed no DRM false recognition.

\begin{figure}[tbp]
  \centering
  \begin{subfigure}[t]{0.31\textwidth}
    \centering
    \includegraphics[width=\linewidth,height=4.2cm,keepaspectratio]{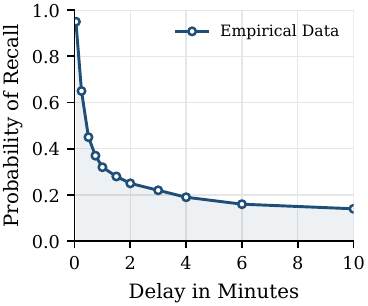}
    \caption{Human memory shows rapid early decay followed by stabilization over time.}
    \label{subfigure1}
  \end{subfigure}\hfill
  \begin{subfigure}[t]{0.31\textwidth}
    \centering
    \includegraphics[width=\linewidth,height=4.2cm,keepaspectratio]{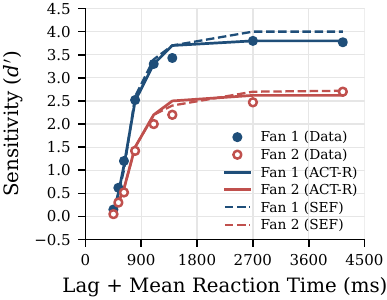}
    \caption{Accuracy of human memory declines as item associations increase. Fan 1 outperforms Fan 2, especially at longer lags.}
    \label{subfigure2}
  \end{subfigure}\hfill
  \begin{subfigure}[t]{0.31\textwidth}
    \centering
    \includegraphics[width=\linewidth,height=4.2cm,keepaspectratio]{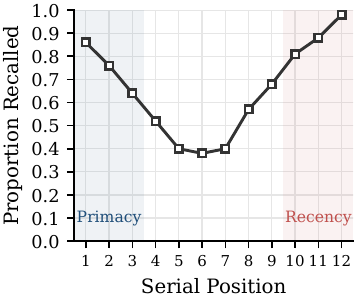}
    \caption{Human recall exhibits a U-shaped serial position curve with primacy and recency effects.}
    \label{subfigure3}
  \end{subfigure}
  \caption{Comparison of human memory patterns with LLM behavior. Data in (a) and (c) are replotted from \citet{rubin1999precise} and \citet{roediger1995creating}; data and model fits in (b) are adapted from \citet{schneider2012modeling}, based on \citet{wickelgren1977associative}.
  (a) Unlike human memory, LLMs do not exhibit intrinsic time-based decay; however, they are limited by a fixed context window.
  (b) LLM recognition accuracy declines under increased associations to an item, reflecting human-like memory patterns, with both LLMs and humans exhibiting the fan effect.
  (c) In 12-item lists, humans show a U-shaped serial position curve in recall and a flatter one in recognition, whereas LLMs recognized studied words perfectly at every position.}
  \label{fig:human_comparison}
\end{figure}

\subsection*{Comparison of LLM and Human Memory Effects}
In this work, we use human memory research as an experimental toolkit to gain insights into the functioning of language models. By aligning established effects of human recall and recognition with LLM behavior, we aim to better understand the extent to which these systems exhibit similar memory patterns. By drawing on canonical experimental paradigms from cognitive psychology, we also compare empirical human data examples with model behavior. \citet{rubin1999precise} showed (Figure~\ref{subfigure1}) that human memory decays in a predictable way over time, with recall probability decreasing sharply within the first minute post-encoding before stabilizing. For instance, according to the list length effect, recall accuracy declines with increasing list length. A comparable decline in the LLM establishes a behavioral parallel. In our task, memory load is manipulated within a constrained context; in humans, interference may also accumulate with elapsed time and intervening experience. Strengthening a selected study fact through repetition did not reduce accuracy for the non-repeated items (Figure~\ref{result}, list strength panel), consistent with the null list-strength effect in human recognition \citep{ratcliff1990list}.

\citet{schneider2012modeling} fit an ACT-R model to speed--accuracy tradeoff data for associative recognition, using the shifted exponential function, a descriptive curve for such data, as a benchmark for the model's fit. In the data of \citet{wickelgren1977associative} that they modeled (Figure~\ref{subfigure2}), fan-1 accuracy exceeds fan-2 accuracy at long lags; that is, higher fan lowers asymptotic accuracy for these briefly studied pairs. Our experiments show a parallel pattern: with less associative interference, LLMs show better discriminability in recognition tasks. This correspondence supports the use of the fan paradigm as a diagnostic of LLM behavior; it does not show that ACT-R associative spreading is implemented inside the model.

\begin{figure}[tbp]
  \centering
  \includegraphics[width=\textwidth,height=0.4\textheight,keepaspectratio]{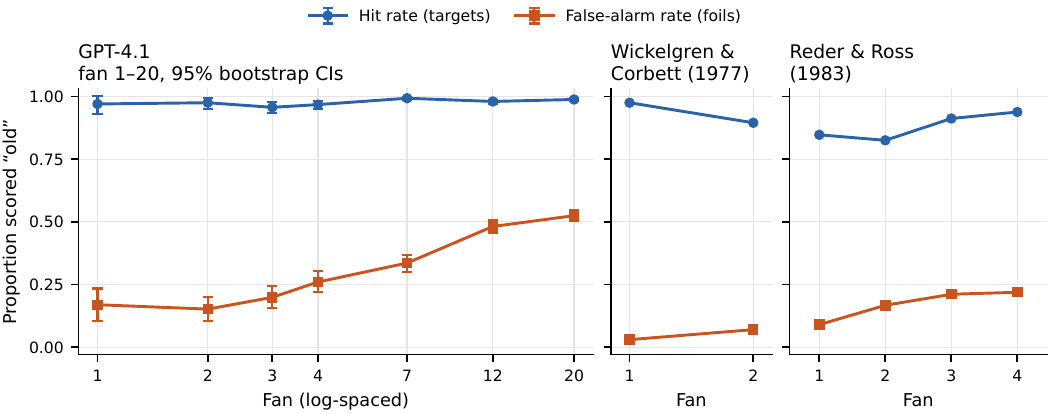}
  \caption{Proportion of probes scored ``old'' by fan, studied items in blue and foils in orange. A probe is scored ``old'' when its log-odds exceeds zero, not by the emitted label. Left: GPT-4.1, fan 1 to 20, 9,800 pooled trials, with 95\% bootstrap confidence intervals from 2,000 resamples. Center: \citet{wickelgren1977associative}, fan 1 to 2, asymptotic. Right: \citet{reder1983integrated}, related foils, fan 1 to 4. Panels share the y-axis only; the two human sources report proportions without trial-level data, so no CIs are shown. The left panel replots the hit and false-alarm rates of Figure~\ref{fig:extended_effects}b for comparison with the human data.}
  \label{fig:fan_llm_human}
\end{figure}

A key divergence concerns serial position. Human free recall typically shows a U-shaped serial position curve \citep{glanzer1966two}, with better memory for the first and last items of a list (Figure~\ref{subfigure3}; \citealp{roediger1995creating}); in recognition, the curve is usually flatter. With the 12-word lists experiment, the model recognized studied words perfectly at every position and showed no serial position effect. The decline in accuracy across the test sequence of the extended-range fan experiment (Figure~\ref{fig:extended_effects}a) should not be read as a primacy effect: hit rates stayed near ceiling throughout, and the decline came from rising false alarms as the context grew. We acknowledge that behavioral similarity does not imply shared mechanisms. Multiple computational models with incompatible assumptions can generate similar or the same memory phenomena \citep{anderson1978arguments,hintzman1988judgments}. Our results characterize behavioral alignment rather than mechanistic equivalence.

Figure~\ref{fig:fan_llm_human} places the extended-range result beside two human data sets. In \citet{reder1983integrated}, as in the model, false alarms rose with fan, from .09 at fan 1 to .22 at fan 4, while hit rates showed no consistent decline. In \citet{wickelgren1977associative}, false alarms also rose, from .03 to .07, but hit rates fell as well, from .98 to .90. The rise in false alarms with fan is therefore common to the model and both human data sets, whereas the stability of the model's hits matches only the second. The comparison concerns the direction of change: the human studies span fan 1--2 and 1--4, the model was tested up to fan 20, and the human values are reported proportions rather than trial-level data.

Overall, our findings, as summarized in Table~\ref{tab:memory_effects}, reveal that language models exhibit human-like patterns in phenomena such as the fan effect, list length/strength effect, and DRM effect. However, their performance appears to be less influenced by factors such as lexicality manipulation and positional effects. Additionally, large language models also exhibit human-similar generalization ability, as they show a high proportion of generalizing critical lure words even when these words are not presented in the study list. An interesting result is that language models tend to exhibit memory errors similar to humans in recognition tasks, yet their performance in recall tasks remains nearly perfect. The extended-range fan experiment refines the recognition results. Rather than attributing the model's performance to a generic memory cache, the response-evidence analysis shows that studied targets remain strongly represented while increasingly familiar recombinations acquire greater ``old'' evidence. This pattern shows competition among contextual associations and familiarity-like evidence aggregation as more specific candidate mechanisms.

\subsection*{Comparison of LLM and Human \texorpdfstring{$d'$}{d'} Values}
\begin{figure}[tbp]
  \centering
  \includegraphics[width=\textwidth,height=0.5\textheight,keepaspectratio]{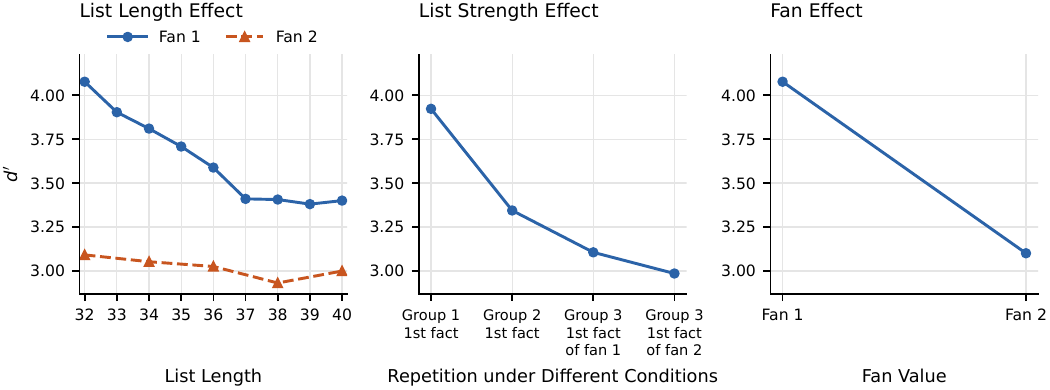}
  \caption{LLM performance across memory effects measured by $d'$. Mean $d'$ values are shown for three experimental groups: (a) memory sensitivity and discriminability decrease as list length increases, particularly under lower fan conditions; (b) recognition sensitivity in the four conditions, which differ in the fan of the study list, so the decline across conditions reflects fan; and (c) fan effect, demonstrating lower overall $d'$ for Fan 2 compared with Fan 1, consistent with classical interference patterns in human associative recognition.}
  \label{figure4}
\end{figure}

For human associative recognition, \citet{schneider2012modeling} estimated the asymptotic sensitivity in the group data of \citet{wickelgren1977associative}, that is, the accuracy reached when participants have ample time to retrieve. The estimates were $d'$ = 3.92 for fan 1 and 2.45 for fan 2; even with unlimited retrieval time, higher fan lowered discriminability (Figure~\ref{subfigure2}). In the classic-range fan experiment (Figure~\ref{figure4}), mean $d'$ was approximately 4.1 for Fan 1 and 3.1 for Fan 2. This reproduces the direction of the human comparison at the behavioral level, but it does not establish a common encoding or retrieval mechanism. The higher $d'$ values for the online LLM indicate stronger recognition sensitivity under these specific task conditions.

Figure~\ref{figure4} also summarizes sensitivity across the list-length and list-strength manipulations in Experiments 1 and 2. Panel (a) shows a decline in $d'$ as the number of studied pairs increases, consistent with the direction of the human list-length effect. Fan 1 consistently outperforms Fan 2 in $d'$ across the tested list lengths, indicating that less association imposes less retrieval interference. Panel (b) shows sensitivity in the four list-strength conditions. Because these conditions differ in fan, the decline across them reflects fan rather than repetition; within each condition, repetition did not reduce sensitivity.

\begin{figure}[tbp]
  \centering
  \includegraphics[width=0.62\textwidth]{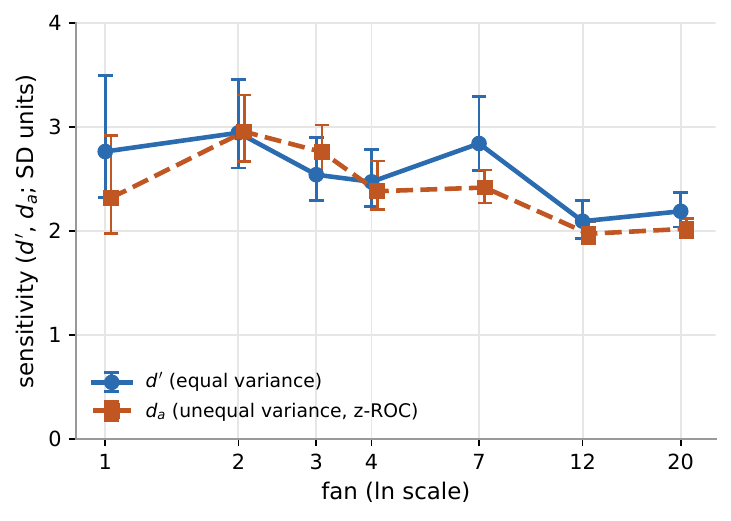}\vspace{-2mm}
  \caption{Target--foil discriminability by fan: $d'$ (blue) and unequal-variance $d_a$ (orange), in standard-deviation units. Error bars show 95\% bootstrap confidence intervals.}
  \label{fan_discriminability}
\end{figure}

The extended-range fan experiment broadened this sensitivity analysis to seven fan levels ranging from 1 to 20. Figure~\ref{fan_discriminability} compares equal-variance $d'$ with the unequal-variance sensitivity index $d_a$, estimated from per-fan z-ROC analyses. Including $d_a$ allows the analysis to accommodate unequal target and foil variances, complementing the equal-variance $d'$ used for the classic-range fan experiment. The two indices differ at some fan levels. Both are lower at fan 20 than at their respective peaks, but neither decreases monotonically with increasing fan. These results qualify the interpretation of the foil evidence shift described above: increasing fan clearly increases false alarms to recombined foils, but the sensitivity estimates do not show a corresponding decline at every successive fan level. Thus, the extended-range experiment reveals a foil-driven change in recognition behavior without a strictly monotonic decline in target--foil discriminability.

\section{Discussion}
The effects we examine here capture core aspects of memory performance. Some focus on associations, while others highlight how order and repetition shape recall. Psychologists studied these effects because each addresses a foundational question: whether memory is limited by capacity (list length), by interference (list strength, fan effect), by temporal order (serial position effect), by meaning (lexicality), by its reconstructive rather than reproductive nature (false memories), or by the tradeoff between preserving detail and enabling abstraction (generalization). Together, these paradigms show the tradeoffs that make memory both adaptive and fallible and explain why they remain benchmarks for understanding human and artificial systems. Our findings show that LLMs echo these tradeoffs, though with different balances, revealing both parallels and divergences in how natural and artificial systems remember. \citet{brainerd2008developmental} show how developmental shifts in false memory reflect the balance between verbatim and gist memory, pointing to next steps for testing whether artificial systems are also susceptible to these tradeoffs.

The extended-range fan experiment refines the classic-range fan result. Higher fan did not mainly reduce recognition of studied targets; instead, it increased the evidence assigned to recombined foils relative to a fixed decision criterion. This pattern is compatible with global-matching accounts of familiarity: a foil contains familiar constituents, and its ``old'' evidence grows as those constituents participate in more studied facts.

There is a reason to expect a model of human memory to have purchase on a language model. The rational analysis of memory begins from the premise that memory does the best job it can of predicting the next piece of information one is going to need \citep{anderson1989human,schooler2017adaptive}. A language model is trained to a target that is formally close, namely to predict the next token given the preceding context.
This correspondence motivates modeling as the next step. In rational analysis, the activation of a memory is interpreted as the log odds, and our evidence score $a(q)$ is likewise a log odds, computed from the model's next-token distribution, so the two can be compared on a common scale. The extended-range data already distinguish between the competing explanations. In ACT-R, fan reduces the associative strength $S_{ji}$ reaching a studied chunk (Equation~\ref{eq:actr_fan}), so spreading-activation accounts predict weaker evidence for targets as fan increases. The model instead kept target evidence high while foil evidence rose toward the criterion. This is what global-matching models predict: because familiarity sums a probe's match to all studied items, it rises with how often the probe's components were studied \citep{clark1996global,hintzman1988judgments}. In future work, we plan to fit an ACT-R activation model and a global-matching model to the trial-level distributions of $a(q)$ across fan, and to compare their fits by likelihood to determine which account better explains the data.

\bibliographystyle{apacite}
\bibliography{sample-base}

\clearpage
\appendix
\titleformat{\section}[display]
  {\normalfont\Large\bfseries}   
  {Appendix}                     
  {0.6ex}                       
  {\thesection\quad}            
\section{Experimental Design}\label{appendix_a}

\subsection*{Common Framework}
Experiments 1--4 used the person--location study-list framework and the three fan conditions defined in the main text. Group 1 contained only fan-1 pairs and Group 2 only fan-2 pairs. In the mixed-fan baseline (Group 3), the study list contained 24 persons and 24 locations; 16 of each appeared once, and 8 of each appeared twice, yielding 16 fan-1 and 16 fan-2 pairs. Across experiments, a \textit{hit} is an \texttt{old} response to a studied item, and a \textit{false alarm} is an \texttt{old} response to a non-studied item. Hit and false-alarm rates are the proportions of such responses among studied and non-studied items, respectively.

\paragraph{Experiment 1: List-length effect.}
Recognition accuracy typically declines as a study list grows longer, because each additional item increases interference at retrieval. To test whether LLMs show this effect, we varied list length within Groups 1 and 2 while keeping all other aspects of the procedure identical to the fan-effect experiment. Starting from the same 32-pair list, Group 1 increased from 32 to 40 pairs in steps of one (32, 33, \ldots, 40), drawn from a pool of 40 fan-1 person--location pairs. Group 2 followed the same procedure but increased length in steps of two pairs.

\paragraph{Experiment 2: List-strength effect.}
The ``strength'' of an item refers to how well it is encoded or how robustly it is represented in memory, and it can be increased by repeated exposure. We therefore manipulated the strength of a single study fact while holding all other facts constant. Within each group, one selected study fact was presented twice or five times in total (one or four additional presentations); all other facts were presented once. The strengthened fact was the first study fact in Groups 1 and 2. In Group 3, two conditions strengthened the first fan-1 fact and the first fan-2 fact, respectively, each at both presentation levels.

\paragraph{Experiment 3: Fan effect.}
We examined fan effects using a controlled-variable design. Because recognition performance also depends on list length (Experiment 1), study-list length was held constant at 32 person--location pairs across conditions, so that differences in performance could be attributed to fan rather than to the amount of studied material. Fixing the study list also fixes the test list, which is twice the length of the study list. Groups 1 and 2 tested fan 1 only and fan 2 only, respectively.

\textit{Extended-range fan experiment.} To examine a much wider range of associative loads and the mechanism of recognition errors, fan was logarithmically spaced at $1, 2, 3, 4, 7, 12,$ and $20$. At each level, 20 object and 20 location identifiers were generated (140 of each per stimulus set), yielding $20, 40, 60, 80, 140, 240,$ and $400$ distinct studied facts, respectively (980 object--location facts per replication). Each fact was presented once, so a fan-$N$ item appeared in $N$ distinct studied facts. Five independent replications were run, each with newly generated nonsense identifiers and study and test lists.

\paragraph{Experiment 4: Lexicality manipulation.}
If LLMs encode words more extensively than nonwords, as people do \citep{xu2007modeling}, replacing person and location words with nonsense strings should lower hits and raise false alarms. Using Group 1, we replaced (a) persons only, (b) locations only, or (c) both, in both the study and test lists. Each replaced word became a random, length-matched string of letters and digits (e.g., \texttt{library} $\rightarrow$ \texttt{s3mc01m}).

\paragraph{Experiment 5: Serial position effect.}
The model studied each of the first 6 associate lists (the first 12 words of each) and was tested immediately after each list. Each 42-word recognition test contained the 12 studied items and 30 non-studied items: (a) 6 critical lures, (b) 12 words unrelated to any list, and (c) 12 words weakly related to the lists (2 per list, from serial positions 14 and 15). We recorded the recognition rate of studied words as a function of serial position.

\paragraph{Experiments 6 and 7: DRM effect and generalization.}
Following \citet{roediger1995creating}, the model studied 120 words from 8 associate lists ($8 \times 15$; Supporting Information Table~S29, which also lists the math problems used in the delayed condition). The 48-word recognition test contained 24 studied words (serial positions 1, 7, and 15 of each studied list) and 24 non-studied words: the critical lures of all 12 lists (8 studied, 4 non-studied) and 12 items from the 4 non-studied lists (positions 1, 7, and 15). The model labeled each word \texttt{old} or \texttt{new}. To examine whether the delay between study and test affects false memory, we included two conditions. In the \textit{immediate} condition, the model took the recognition test directly after studying the list. In the \textit{delayed} condition, the model first completed a series of simple math problems, which served as a filler task, before taking the test. The DRM effect was measured by hit and false-alarm rates. Generalization was indexed by false alarms to critical lures of studied lists: endorsing these lures indicates that the model generalized from their associates, although the lures were never presented.

\end{document}